# Practical Formula for Laser Intensity at Beam Spray Onset

Pavel M. Lushnikov[*,#,&] and Harvey A. Rose[&]

[*] Landau Institute for Theoretical Physics,
Kosygin St. 2, Moscow, 119334, Russia
[#] Department of Mathematics and Statistics,
University of New Mexico, Albuquerque, NM 87131, USA
[&] Theoretical Division, Los Alamos National Laboratory,
MS-B213, Los Alamos, New Mexico, 87545, USA

A practical formula for calculating laser intensity at the onset of beam spray is presented. This result is based upon our previously published work[1,2].

## I.     INTRODUCTION

In laser fusion, typical laser beam intensities are so large that self-focusing in plasma may lead to unacceptably large changes in beam angular divergence, "beam spray". When strong temporal beam smoothing is used, classical self-focusing estimates[3], based on static hydrodynamic response models, are not relevant. Instead the plasma responds dynamically: ion-acoustic waves are generated whose dynamic response may exceed the static response by as much as two orders of magnitude. A practical formula is presented below, suitable for use in Inertial Confinement Fusion design codes, for the laser beam intensity at onset of beam spray, taking into account plasma dynamics. The onset intensity rapidly decreases with addition of high ionization state "dopant", *e.g.*, addition of small amounts of xenon to hydrogen plasma. The dynamic beam spray onset intensity value is typically small compared to classical self-focusing estimates.

## II. THE PRACTICAL FORMULA

If strong temporal beam smoothing is used, *i.e.*, the laser coherence time, $T_c$, is short enough to suppress speckle self-focusing, $c_s T_c < F\lambda_0$, then the laser intensity at beam spray onset is given by

$$\frac{I_{onset}}{W/cm^2} \approx 2.7\times 10^{15} \left(\frac{T_e}{keV}\right)\left(\frac{\mu m}{\lambda_0}\right)^2 \frac{v_{ia}}{F^2}\frac{n_c}{n_e} \mathrm{Re}\left\{(2.2-0.31i)\left[1+\frac{Z^*\left(\frac{k_0\lambda_e}{4F}\right)^{-2/3}}{1-i2.7\sqrt{Z^*\phi}\left(\frac{k_0\lambda_e}{4F}\right)^{1/3}\frac{c_s}{v_e}}\right]^{-1}\right\} \quad (1)$$

Physically this corresponds to gain exponent unity over a speckle length, $7F^2\lambda_0$. If the electron temperature, $T_e$, varies too rapidly in a certain spatial region, $7F^2\lambda_0(\partial T_e/\partial z)/T_e > v_{ia}$, then this result is not applicable in that region.

## III. CONCLUSION

Since laser-plasma-interaction physics is essentially nonlinear in actual plasma experiments, it is not possible to model laser beam propagation independently of global plasma hydrodynamics. So our purpose here is limited to calculation of beam spray onset for *given* plasma parameters such as electron density and temperature. If, in a plasma sub-volume near the region where the laser beam enters plasma, the laser intensity is above the beam spray onset value, then beam propagation over a large fraction of the plasma volume will depart from its classical (*i.e.* geometric optic) trajectory. In such case, the "given" plasma parameters are not reliable. Conversely, if beam spray occurs only near the end of a beam's trajectory (*e.g.*, in the strong absorption region near the hohlraum wall), then beam spray is ignorable.

## IV. SYMBOL DEFINITIONS

$I_{onset}$: laser intensity at beam spray onset

$T_e$: electron temperature

$\lambda_0$: laser wavelength

$\nu_{ia}$: ratio of ion acoustic amplitude damping rate to ion acoustic frequency

$F$: optic f/#, the ratio of optic focal length to lens diameter

$n_e$: electron density

$n_c$: critical electron density for laser light

$k_0$: $k_0 = 2\pi/\lambda_0$

$\lambda_e$: $\lambda_e = \frac{1}{3}\lambda_{ei}\sqrt{\frac{2Z^*}{\pi\phi}}$ with $\lambda_{ei}$ the electron-ion mean free path

$\phi$: $\phi = \frac{4.2 + Z^*}{0.24 + Z^*}$

$c_s$: ion acoustic wave speed

$v_e$: electron thermal speed

$Z^*$: effective ionization, $Z^* = \frac{\sum_i n_i Z_i^2}{\sum_i n_i Z_i}$

---